\documentclass[showpacs,amsmath,amssymb,aps,prb,10pt,
reprint,superscriptaddress]{revtex4-2}
\usepackage{bm}
\usepackage[breaklinks=true,colorlinks=true,linkcolor=blue,urlcolor=blue,citecolor=blue]{hyperref}
\usepackage{dcolumn}
\usepackage{amsmath,amssymb}
\usepackage{mathdots}
\usepackage{natbib}
\usepackage{soul,color}
\usepackage{graphicx}
\usepackage[note-name]{notes2bib}
\usepackage{mathptmx}
\usepackage[margin=1in]{geometry}
\begin{document}

\title{Vortex rivers and multiple voltage transitions in superconducting MgB$_2$ thin films}

\author{Anton Pokusinskyi}
    \email[Corresponding author: ]{anton.pokusinskyi@tu-braunschweig.de}
    \affiliation{Cryogenic Quantum Electronics, Institute for Electrical Measurement Science and Fundamental Electrical Engineering (EMG) and Laboratory for Emerging Nanometrology (LENA), Technische Universit\"at Braunschweig, 38106, Braunschweig, Germany}

\author{Clemens Schmid}
    \affiliation{Faculty of Physics, University of Vienna, Austria}
    \affiliation{Vienna Doctoral School in Physics, Vienna, Austria}

\author{Thomas Hauet}
    \affiliation{Institute Jean Lamour, Université de Lorraine-CNRS, Nancy, France}

\author{Oleksandr Dobrovolskiy}
    \affiliation{Cryogenic Quantum Electronics, Institute for Electrical Measurement Science and Fundamental Electrical Engineering (EMG) and Laboratory for Emerging Nanometrology (LENA), Technische Universit\"at Braunschweig, 38106, Braunschweig, Germany}
    \affiliation{FLUXONICS---The European Foundry for Superconducting Electronics e.V., 38116 Braunschweig, Germany}

\date{\today}

\begin{abstract}
Vortex dynamics govern the dissipation and magneto-resistive properties of type-II superconductors. At large transport currents, the current-voltage ($I$-$V$) characteristics of a superconductor in the mixed state usually exhibit a nonlinear upturn followed by an abrupt jump occurring due to a flux-flow instability. However, other transition behaviors are also possible, including the formation of phase-slip lines and normal domains. Which mechanism dominates depends on the sample uniformity, its dimensions relative to the coherence length and penetration depth, and the rates of electron energy relaxation and heat removal. Here, based on the time-dependent Ginzburg-Landau equation, we present the results of numerical modeling of the $I$-$V$ curves of superconducting films with various types of disorder. For a grain-boundary defect mesh, we find multiple voltage transitions overlaid with a nonlinear upturn of the $I$-$V$ curves. For randomly arranged elongated defects, the $I$-$V$ curves exhibit voltage steps, whereas for L-shaped defects oriented perpendicular to the transport current, the $I$-$V$ curves show extended linear regimes separated by voltage transitions. We analyze the evolution of the order parameter along the $I$-$V$ curves and discuss the experimental accessibility of the revealed dynamics regimes.
\end{abstract}


\maketitle

\section{Introduction}

Most superconductors relevant for applications are type-II superconductors\cite{Bra95rpp,Dob26sst}. In magnetic fields between the lower and upper critical values, they are penetrated by Abrikosov vortices \cite{Bra95rpp}. The vortex-vortex, vortex-current, vortex-edge, and vortex-defect interactions \cite{Vod19sst,Kog20prb,Pat21prb,Kog22prb,Kar24pra} determine both the arrangement of vortices and their dynamics under an applied transport current. While the enhancement of vortex pinning -- i.e., the spatial anchoring of fluxons to structural imperfections -- is crucial for improving a superconductor's capacity to carry dissipation-free currents \cite{Rui26pms}, further phenomena are essential for the current-driven breakdown of superconductivity at high vortex velocities \cite{Dob24inb}. Namely, at sufficiently large transport currents, the current–voltage ($I$-$V$) characteristics of a superconductor are expected to exhibit a nonlinear upturn, signaling the regime of nonlinear conductivity \cite{Lar76etp}. In general, one would expect a continuous transition from the low-resistive flux-flow regime to the high-resistive state with increasing transport current. However, the flux-flow regime is typically destroyed abruptly, resulting in a sudden voltage jump due to flux-flow instabilities (FFIs) \cite{Mus80etp,Dmi81phb,Kle85ltp}. 

Close to the critical temperature $T_\mathrm{c}$, FFI arises from the escape of quasiparticles from the vortex cores, leading to vortex shrinkage and a reduction in vortex viscosity, followed by further vortex acceleration. This Larkin-Ovchinnikov mechanism~\cite{Lar76etp} is rooted in a \emph{non-thermal} electron distribution function. By contrast, at $T \ll T_\mathrm{c}$, FFI is of \emph{thermal} origin and is described by the Kunchur hot-electron model~\cite{Kun02prl}, in which the electron temperature $T_\mathrm{e}$ is elevated by dissipation. The resulting increase in $T_\mathrm{e}$ generates additional quasiparticles, expands the vortex core, and reduces viscous drag due to a softening of vortex-profile gradients.

The nonuniform current distribution~\cite{Emb17nac}, pinning landscape~\cite{Sil12njp,Bud22pra}, metallic capping layer~\cite{Per02pcs,Att12pcm}, and heat removal rate all influence the nonequilibrium state associated with high-velocity vortex dynamics~\cite{Bez19prb}. Furthermore, the $I$-$V$ characteristics may exhibit negative differential resistance~\cite{Kun01prl,Ust24prb} and multiple voltage transitions~\cite{Mis07prb,Gut09prb,Ada15prb}. As the vortex velocity increases, the unpaired electrons escaping from the vortices lead to a suppression of the order parameter along the vortex trajectories. The track left by a moving vortex attracts additional vortices, leading to local overheating \cite{Bez19prb}. If heat removal from the superconductor is efficient enough, the trajectories of fast-moving vortices coalesce into \emph{vortex rivers}~\cite{Sil10pcs,Vod19sst}. Vortex rivers are essentially phase-slip lines \cite{Dmi06sst,Emb17nac} along which the superconducting order parameter is suppressed. In narrow strips, phase slips can also occur in the absence of vortices~\cite{Siv03prl}. In the following, we use the term ``vortex river'' to emphasize that the considered phase-slip lines are formed by fast vortex motion, irrespective of whether the vortices are induced by an external magnetic field~\cite{Sil10pcs,Dob20nac} or arise from the self-field of the transport current~\cite{Bez22prb,Bev23pra,Ust24prb}. Finally, if heat removal is insufficient~\cite{Bez84ltp}, vortex rivers evolve into normal domains that grow until the entire sample transitions to the normal-conducting state~\cite{Vod19sst}.

Recently, we experimentally studied current-driven transitions to the normal state in hybrid structures based on MgB$_2$ superconducting films~\cite{Gru24msc,Sch26prb}. The films differ in thickness, structure, film-buffer interface, and capping layer. Figure~\ref{fig:Exp} shows representative $I$-$V$ curves for the three studied stacks, exhibiting a variety of behaviors, ranging from distinct voltage steps (Fig.~\ref{fig:Exp}a) to multiple voltage transitions superimposed on a nonlinear upturn (Fig.~\ref{fig:Exp}b), as well as
multiple voltage jumps between the linear sections of the $I$-$V$ characteristics (Fig.~\ref{fig:Exp}c). A systematic analysis of the experimental data is the subject of dedicated publications.

Previously, vortex dynamics has been extensively studied in the presence of grain boundaries~\cite{Rei00prb}, random~\cite{Pat21prb} and periodic~\cite{Rei97prl} pinning, as well as in confined geometries~\cite{Ber12prl,Emb17nac}, among others. 

Here, we present numerical modeling of the $I$-$V$ characteristics of superconducting films based on the time-dependent Ginzburg–Landau (TDGL) equation. Specifically, we consider defect configurations mimicking grain boundaries and elongated defects of two distinct geometries. For randomly arranged elongated defects, the $I$-$V$ curves exhibit voltage steps. For a grain-boundary mesh, we find multiple voltage steps superimposed on a nonlinear upturn of the $I$-$V$ curve. For L-shaped defects oriented perpendicular to the transport current, the $I$-$V$ curves show extended linear regimes separated by voltage transitions. We analyze the evolution of the order parameter along the $I$-$V$ characteristics and discuss the accessibility of the observed regimes in experiments.

\section{TDGL modeling}
\begin{figure}
    \centering
    \includegraphics[width=6.5cm]{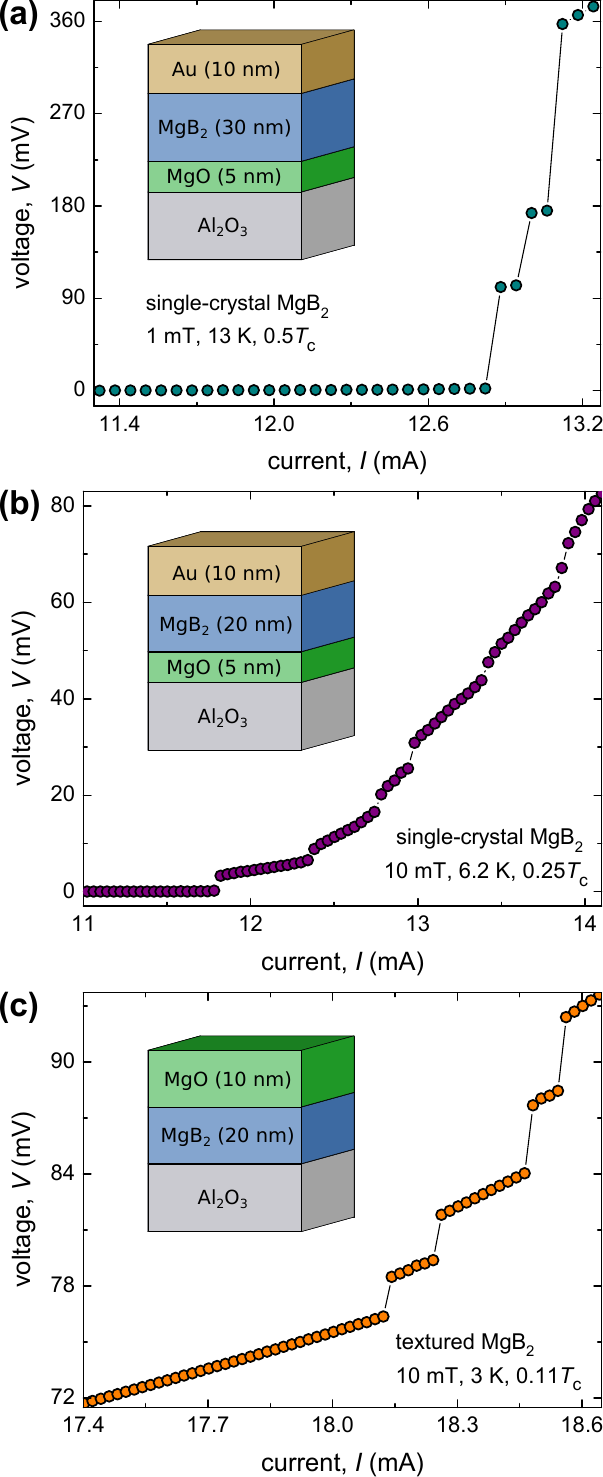}
    \caption{Representative experimental $I$-$V$ curves for MgB$_2$-based hybrid structures after Ref.~\cite{Sch26prb}: (a) the current-driven resistive transition is mediated by voltage steps; (b) multiple voltage transitions are superimposed on a nonlinear $I$-$V$ curve; (c) multiple voltage steps occur within the linear sections of the $I$-$V$ curve.}
    \label{fig:Exp}
\end{figure}

The primary objective of the numerical modeling is to assess how different defect configurations affect the $I$-$V$ curves of superconducting strips. The experimental dependences in Fig.~\ref{fig:Exp} serve only as representative examples of observed $I$-$V$ shapes and are not intended for quantitative reproduction. Moreover, the modeling does not account for proximity effects~\cite{Pam94prb,Per02pcs,Kom14apl} at the MgB$_2$/Au interface and the two-band nature of superconductivity in MgB$_2$~\cite{Nag01nat,Tsu03prl,Pfa24apl,Pok24ltp,Pok25rrl}. Accordingly, the discussed results are not specific to MgB$_2$ but instead capture generic vortex dynamics in the presence of structural disorder.

In two dimensions, as appropriate for thin-film geometries, the generalized TDGL equation reads \cite{Kra78prl, Bis23cpc}
\begin{equation}
    \begin{split}
    \frac{u}{\sqrt{1+\gamma^2|\psi|^2}}&\left(\frac{\partial}{\partial t}+i\varphi+\frac{\gamma^2}{2}\frac{\partial |\psi|^2}{\partial t}\right)\psi\\
    &= (\nabla-i\mathbf{A})^2\psi + (\epsilon-|\psi|^2)\psi
    ,
    \end{split}
    \label{eq:tdgl_psi}
\end{equation}
where $\psi(\mathbf{r},t) = |\psi| e^{i\theta}$ is the superconducting order parameter with phase $\theta$. $u = \pi^4 / 14\zeta(3) \approx 5.79$ denotes the ratio of the relaxation times of its amplitude and phase in the dirty limit and $\zeta(x)$ is the Riemann zeta function. The parameter $\gamma = 2\tau_\mathrm{E} \Delta_0 / \hbar$ depends on the inelastic electron-phonon scattering time $\tau_\mathrm{E}$ and the zero-field superconducting gap $\Delta_0$. $\varphi(\mathbf{r}, t)$ is the scalar electric potential and $\mathbf{A}$ the vector potential. The real-valued function $\epsilon(\mathbf{r}) \in [-1,1]$ locally modulates the order parameter and models its suppression at defects~\cite{Kos16prb}. The defect strength is parametrized as $\epsilon = 1 - \Delta\epsilon$, so that larger $\Delta\epsilon$ corresponds to stronger order-parameter suppression.

The total current density consists of the supercurrent density
$\mathbf{j}_\mathrm{s} = \mathrm{Im}\!\left[\psi^\ast(\nabla - i\mathbf{A})\psi\right]$
and the normal current density
$\mathbf{j}_\mathrm{n} = -\nabla\varphi - \partial \mathbf{A}/\partial t$. The continuity condition for the total current, $\nabla \cdot (\mathbf{j}_\mathrm{s} + \mathbf{j}_\mathrm{n}) = 0$, implies that the scalar potential $\varphi(\mathbf{r}, t)$ satisfies a Poisson equation
\begin{equation}
    \nabla^2 \varphi
    = \nabla \cdot \mathrm{Im}[\psi^\ast(\nabla - i\mathbf{A})\psi]
    - \nabla \cdot \frac{\partial \mathbf{A}}{\partial t}.
    \label{eq:tdgl_poisson}
\end{equation}

\begin{figure*}
    \centering
    \includegraphics[width=\linewidth]{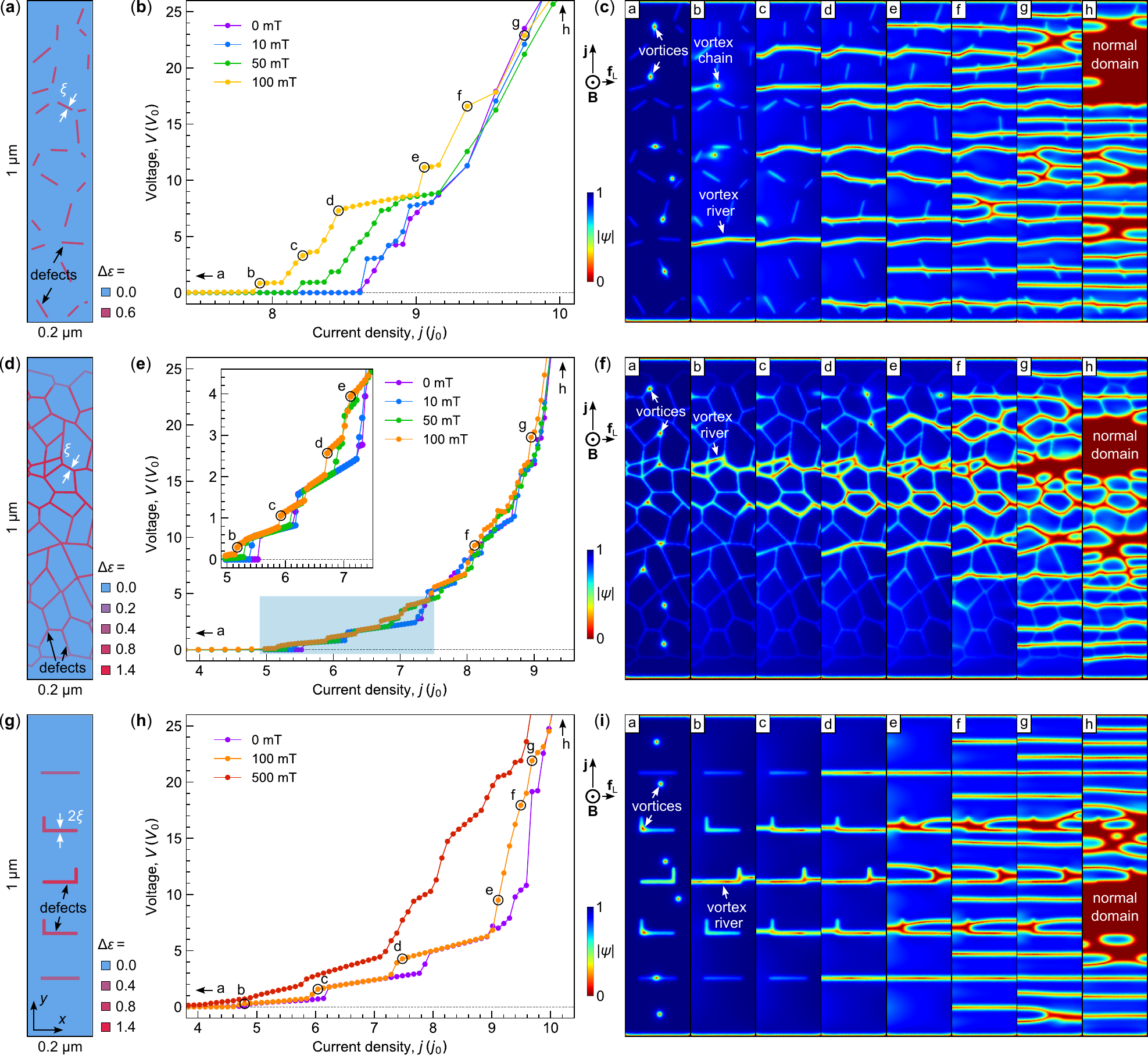}
    \caption{TDGL modeling results for three types of defect structures: linearly-extended defects (a-c), grain-boundary defects mesh (d–f), and L-shaped defects (g–i). Panels (a), (d), and (g) show the geometries of the superconducting (SC) strips, indicating the spatial distribution and strength of order-parameter suppression used to model the defects. The calculated current–voltage ($I$–$V$) characteristics are presented in panels (b), (e), and (h) for the respective structures. Labeled points in panels (b), (e), and (h) correspond to representative dynamical states, with snapshots of the order-parameter magnitude $|\psi|$ shown in panels (c), (f), and (i), respectively. The current density $j$ is expressed in units of $j_0 = 4 \times 10^{-2}\xi B_\mathrm{c2}/(\mu_0 \lambda^2)$, where $B_\mathrm{c2} = \Phi_0/(2\pi \xi^2)$ is the upper critical field. The voltage $V$ is given in units of $V_0 = \xi j_0/\sigma$, where $\sigma$ denotes the normal-state conductivity.}
    \label{fig:modeling_I-V}
\end{figure*}

In Eqs.\,\eqref{eq:tdgl_psi} and \eqref{eq:tdgl_poisson}, the order-parameter magnitude is scaled to the superconducting carrier density, $|\psi|^2 = n_s$. Lengths are measured in units of the coherence length $\xi$, and time is scaled by $\tau_0 = \mu_0 \sigma \lambda^2$, where $\sigma$ is the normal-state conductivity and $\lambda$ the London penetration depth. The vector potential is scaled by $\xi B_\mathrm{c2}$, where $B_\mathrm{c2} = \Phi_0 / (2\pi \xi^2)$ is the upper critical field and $\Phi_0 = h / (2e)$ the magnetic flux quantum. The external current density $j_\mathrm{ext}$ and electric potential $\varphi$ are scaled by $4 \xi B_\mathrm{c2} / (\mu_0 \lambda^2)$ and $\xi j_0 / \sigma$, respectively.

At superconductor--vacuum interfaces, where vortices enter or exit the strip, the boundary conditions are
$\mathbf{n}\cdot(\nabla - i\mathbf{A})\psi = 0$ and $\mathbf{n}\cdot\nabla\varphi = 0$, where $\mathbf{n}$ is the outward unit normal vector. At superconductor--normal-metal interfaces supplying the transport current $j_\mathrm{ext}$, the boundary conditions are $\psi = 0$ and $\mathbf{n}\cdot\nabla\varphi = j_\mathrm{ext}$. TDGL simulations are performed for $200\times 1000\,\mathrm{nm}$ (width $\times$ length) strips for a range of applied magnetic fields. The maximum mesh size is set to $\xi/2$. The material parameters used are summarized in Table~\ref{tab:tdgl_material_params}. 

\begin{table}[h!]
\caption{Material parameters used in the simulations.}
\label{tab:tdgl_material_params}
\small
\begin{center}\begin{tabular}{lcc}
 \hline
 \textbf{Parameter} & \textbf{Denotation} & \textbf{Value}\\
 \hline
 Clean-limit coherence length & $\xi_0$ & 9 nm
 \\
 Electron mean free path & $l$ & 3 nm
 \\
 Coherence length & $\xi(0) = 0.85 \, \sqrt{\xi_0 l}$ & 5 nm
 \\
 London penetration depth & $\lambda(0)$ & 100 nm
 \\
 Inelastic scattering coefficient & $\gamma$ & 10
 \\
 Relative temperature & $T/T_\mathrm{c}$ & 0.3
 \\
 Coherence length & $\xi=\frac{\xi(0)}{\sqrt{1-T/T_\mathrm{c}}}$ & 5.2\,nm
 \\
 Penetration depth & $\lambda= \frac{0.615 \, \lambda(0)\sqrt{\xi_0}}{\sqrt{l (1-T/T_\mathrm{c})}}$ & 127\,nm\\
\hline
\end{tabular}\end{center}
\end{table}

\section{Results and Discussion}
Figure~\ref{fig:modeling_I-V} presents TDGL modeling results for three different types of defects in a superconducting strip. Panels (a,d,g) show defect distributions and geometries, (b,e,h) the $I$-$V$ curves, and (c,f,i) snapshots of the magnitude of the superconducting order parameter.

In panel (a) of Fig.~\ref{fig:modeling_I-V} we consider linearly extended defects of width $\simeq \xi$ and length of $2\xi-15\xi$, which are not interconnected and are randomly oriented with respect to the transport current direction. Within the defects, the superconducting order parameter is suppressed by 30\%, and the defects are positioned such that they are at least $\sim 3\xi$ away from the strip edges. 
This is done to minimize suppression of the edge barrier and to prevent vortex nucleation from being confined to a small number of spurious locations along the strip edge.

In panel (d) of Fig.~\ref{fig:modeling_I-V} we examine a grain-boundary defect mesh in which the linearly extended defects are interconnected and reach the strip edges. This arrangement of defects leads to multiple edge locations where the barrier is suppressed and vortex entry is facilitated. The defects are of width $\simeq \xi$ and variable length. The order parameter within the defects is suppressed by $20\%$-$70\%$. 

In panel (g) of Fig.~\ref{fig:modeling_I-V}, we consider L-shaped defects placed at equal intervals along the $y$-direction, with their long segments oriented perpendicular to the transport current direction. The orientation of neighboring defects is alternated by mirroring each defect with respect to the $y$-axis to suppress nonreciprocity along the $x$ direction. The outer defects are modeled as straight to avoid symmetry breaking near the superconductor--normal lead boundaries. The defects have a width of $\simeq 2 \xi$, lateral dimensions of $110 \times 50$\,nm$^2$, and a spatial suppression of the order parameter of $20\%$–$70\%$.

Figures~\ref{fig:modeling_I-V}(b,e,h) show the simulated $I$-$V$ curves. In all cases, the curves exhibit a zero-voltage plateau at sub-depinning currents and a nonlinear resistive regime with multiple voltage transitions at higher currents. The behavior depends strongly on the defect configuration: randomly arranged elongated defects produce distinct voltage steps; the grain-boundary defect mesh yields multiple voltage transitions with a nonlinear upturn of the $I$-$V$ curve; and L-shaped defects give rise to extended linear sections separated by voltage transitions.

Figures~\ref{fig:modeling_I-V}(c,f,i) show instantaneous snapshots of the magnitude of the superconducting order parameter at points a--h along the respective $I$-$V$ curves at 100\,mT. This field corresponds to a moderate vortex density (about 10 vortices in the strip in the ideal equilibrium configuration), with an intervortex spacing of $a \simeq 150$\,nm. At sub-depinning currents, most vortices are pinned at defects in all cases, with a smaller number occupying interstitial positions. We now discuss the sequence of dynamical states for each case individually.

For randomly oriented, disconnected linear defects, the first resistive state corresponds to the formation of a vortex chain, which evolves along the most linear sequence of defects into a vortex river [regime b in panel (c) of Fig.~\ref{fig:modeling_I-V}]. With increasing current, vortex chains are accelerated along additional defect sequences [regimes c–e in panel (c)], giving rise to voltage steps in the $I$-$V$ curve. At higher currents, vortex rivers coalesce into normal domains that grow until the entire sample transitions to the normal conducting state [regimes f–h in panel (c)].

In contrast, for the grain-boundary mesh the trajectories of moving vortices are confined to the grain boundaries [regimes b–d in panel (f) of Fig.~\ref{fig:modeling_I-V}]. The simulations indicate that these boundaries, along which the order parameter is suppressed, channel a significantly larger number of vortices than in the case of disconnected defects, where new vortex paths continuously emerge [regimes b–d in panel (c)]. This behavior is attributed to multiple edge locations with reduced barrier strength. With increasing transport current, the branched regions of suppressed order parameter expand and coalesce, ultimately driving the entire strip into the normal conducting state [regimes e–h in panel (f)].

Increasing the transport current in the case of L-shaped defects leads to the disappearance of interstitial vortices and the formation of vortex rivers along the L-shaped defects [regimes b–d in panel (i) of Fig.~\ref{fig:modeling_I-V}]. With further increase in current, vortices begin to enter at arbitrary locations along the strip edge, and the vortex rivers, as in the other cases, evolve into normal domains [regimes e–h in panel (i)]. In this way, randomly arranged elongated defects produce voltage steps in the $I$-$V$ curves. For a grain-boundary mesh, multiple voltage steps appear on top of a nonlinear upturn, while L-shaped defects yield linear regimes separated by voltage transitions.

We now discuss the experimental accessibility of the predicted regimes. The fact that the superconducting state does not break down at a single current value but persists over a finite current range requires efficient dissipation of the generated heat~\cite{Bez84ltp}. The observation of extended dissipative regimes in experimental $I$-$V$ curves (Fig.~\ref{fig:Exp} and~\cite{Gru24msc,Sch26prb}) indicates that such heat removal is feasible. A more accurate TDGL description requires solving a  heat-balance equation, e.g., as in Ref.~\cite{Dob20nac}, but this substantially increases computational cost.

Another expected discrepancy between modeling and experiment arises from the difference in system dimensions. The simulated strips ($200\times1000$\,nm$^2$) are much smaller than those used experimentally ($10\times28\,\mu$m$^2$)~\cite{Gru24msc,Sch26prb}. Increasing the system size in simulations would incur substantial computational cost. Qualitatively, the larger experimental strips imply that voltage steps may become smeared due to the much larger number of vortex rivers formed between the voltage leads. At the same time, the smaller strip width hinders vortex penetration at lower magnetic fields~\cite{Plo01prb,Dob20nac}. Accordingly, the features discussed here for narrow strips at $100$\,mT are expected to occur at significantly lower fields in wider experimental films~\cite{Gru24msc,Sch26prb}.

Finally, while the TDGL modeling uses parameters corresponding to MgB$_2$, the discussed results are not specific to this material and are expected to apply to other superconductors as well. The considered defect configurations are experimentally viable and can be realized by various means, for example through the choice of buffer layer~\cite{Sch26prb}, grain coarsening during annealing~\cite{Mak26prb}, or surface patterning with artificial pinning structures~\cite{Dob15met}.

\section{Conclusion}
In summary, based on the time-dependent Ginzburg-Landau equation, we numerically studied the current-voltage characteristics of superconducting films with different types of structural disorder. The simulations show that defect geometry critically affects vortex dynamics and the resulting resistive response. Specifically, a grain-boundary defect mesh yields multiple voltage transitions with a pronounced nonlinear upturn of the $I$-$V$ curves, randomly arranged elongated defects produce distinct voltage steps, and L-shaped defects oriented perpendicular to the current give rise to extended linear regimes separated by voltage transitions. Analysis of the order-parameter dynamics reveals a direct link between defect topology, vortex channel formation, and dissipative regimes. With increasing current, vortex entry becomes progressively less confined, leading to vortex chains that evolve into vortex rivers and eventually turn into expanding normal domains. These results provide a unified picture of how engineered disorder affects vortex dynamics and are relevant for interpreting resistive states in superconducting films.
\vspace{3mm}

\section*{Data availability statement}
The data supporting the findings of this study are openly
available \cite{Pok26mendeley}.

\begin{acknowledgments}
The authors appreciate fruitful discussions with Alexander Kasatkin, Markus Gruber, Corentin Pfaff, Theo Courtois, Karine Dumesnil, and Stephane Mangin. The work of A.P. was funded by the Deutsche Forschungsgemeinschaft (DFG, German Research Foundation) under Germany's Excellence Strategy -- EXC-2123 QuantumFrontiers -- 390837967, project Q-53. A.P. gratefully acknowledges the use of the CryoCore simulation workstation at CryoQuant/TU Braunschweig. C.S. acknowledges financial support by the Vienna Doctoral School in Physics (VDSP). T.H. thanks CCDaum, CC3M and CCMinalor at Institut Jean Lamour for their work on UHV growth, HRTEM imaging and on film patterning  respectively. This work was supported by the French National Research Agency through the France 2030 government grants ``PEPR-SPIN'' ANR-24-EXSP-0012 and ``Lorraine Initiative of Excellence'' ANR-15-IDEX-04-LUE. 
The research is based upon work from COST Action CA21144 (SuperQuMap) supported by the European Cooperation in Science and Technology. This research is funded in part by the Austrian Science Fund (FWF), Grant No. I 6079 (FluMag). For the purpose of open access, the authors have applied a CC BY public copyright license to any Author Accepted Manuscript version arising from this submission.
\end{acknowledgments}

%\section*{Appendix}

\bibliographystyle{apsrev4-2}
\bibliography{./main.bib}

@article{Kra78prl,
title = {Theory of Dissipative Current-Carrying States in Superconducting Filaments},
author = {
Kramer, L. and 
Watts-Tobin, R. J.},
j1 = {PRL},
journal2 = {Physical Review Letters},
journal = {Phys. Rev. Lett.},
publisher = {American Physical Society},
volume = {40},
number = {15},
year = {1978},
month = {04},
pages = {1041--1044},
doi = {10.1103/PhysRevLett.40.1041},

date = {1978/04/10/}
}

@ARTICLE{Dob20nac,
  author = {Dobrovolskiy, O. V. and Vodolazov, D. Yu and Porrati, F. and Sachser,
	R. and Bevz, V. M. and Mikhailov, M. Yu and Chumak, A. V. and Huth,
	M.},
  title = {Ultra-fast vortex motion in a direct-write {Nb-C} superconductor},
  journal = {Nat. Commun.},
  year = {2020},
  volume = {11},
  pages = {3291},
  number = {1},
  day = {03},
  doi = {10.1038/s41467-020-16987-y},
  issn = {2041-1723},
  url = {https://doi.org/10.1038/s41467-020-16987-y}
}

@ARTICLE{Dob15met,
  author = {Dobrovolskiy, O. V. and Huth, M. and Shklovskij, V. A.},
  title = {Alternating current-driven microwave loss modulation in a fluxonic
	metamaterial},
  journal = {Appl. Phys. Lett.},
  year = {2015},
  volume = {107},
  pages = {162603--1-5},
  doi = {http://dx.doi.org/10.1063/1.4934487},
  url = {http://scitation.aip.org/content/aip/journal/apl/107/16/10.1063/1.4934487}
}

@INBOOK{Dob24inb,
  chapter = {9},
  pages = {735--754},
  title = {Fast Dynamics of Vortices in Superconductors},
  publisher = {Elsevier},
  year = {2024},
  editor = {Fomin, V.M.},
  author = {Dobrovolskiy, O.V.},
  series = {Encyclopedia of Condensed Matter Physics, 2nd ed.},
  doi = {10.1016/B978-0-323-90800-9.00015-9},
  url = {https://doi.org/10.1016/B978-0-323-90800-9.00015-9}
}

@article{Ber12prl,
  title = {Large Magnetoresistance Oscillations in Mesoscopic Superconductors due to Current-Excited Moving Vortices},
  author = {Berdiyorov, G. R. and Milo\ifmmode \check{s}\else \v{s}\fi{}evi\ifmmode \acute{c}\else \'{c}\fi{}, M. V. and Latimer, M. L. and Xiao, Z. L. and Kwok, W. K. and Peeters, F. M.},
  journal = {Phys. Rev. Lett.},
  volume = {109},
  issue = {5},
  pages = {057004},
  numpages = {6},
  year = {2012},
  month = {Jul},
  publisher = {American Physical Society},
  doi = {10.1103/PhysRevLett.109.057004},
  url = {https://link.aps.org/doi/10.1103/PhysRevLett.109.057004}
}

@article{Rei97prl,
  title = {Dynamic Phases of Vortices in Superconductors with Periodic Pinning},
  author = {Reichhardt, C. and Olson, C. J. and Nori, Franco},
  journal = {Phys. Rev. Lett.},
  volume = {78},
  issue = {13},
  pages = {2648--2651},
  numpages = {0},
  year = {1997},
  month = {Mar},
  publisher = {American Physical Society},
  doi = {10.1103/PhysRevLett.78.2648},
  url = {https://link.aps.org/doi/10.1103/PhysRevLett.78.2648}
}

@article{Rei00prb,
  title = {Dynamic vortex phases and pinning in superconductors with twin boundaries},
  author = {Reichhardt, C. and Olson, C. J. and Nori, Franco},
  journal = {Phys. Rev. B},
  volume = {61},
  issue = {5},
  pages = {3665--3671},
  numpages = {0},
  year = {2000},
  month = {Feb},
  publisher = {American Physical Society},
  doi = {10.1103/PhysRevB.61.3665},
  url = {https://link.aps.org/doi/10.1103/PhysRevB.61.3665}
}

@article{Mak26prb,
  author = {
    Makhdoumi Kakhaki, Zahra and 
    Pokusinskyi, Anton O. and 
    Avitabile, Francesco and 
    Kumar, Abhishek and 
    Colangelo, Francesco and 
    Cirillo, Carla and 
    Attanasio, Carmine and 
    Dobrovolskiy, Oleksandr V.
    },
  title = {Annealing-induced grain coarsening and voltage kinks in superconducting {NbRe} films},
  journal = {Phys. Rev. B},
  volume = {114},
  issue = {9},
  pages = {094209},
  numpages = {12},
  year = {2026},
  month = {Aug},
  publisher = {American Physical Society},
  doi = {10.1103/hhkw-ccv8},
  url = {https://link.aps.org/doi/10.1103/hhkw-ccv8}
}

@ARTICLE{Plo01prb,
  author = {Plourde, B. L. T. and Van Harlingen, D. J. and Vodolazov, D. Yu.
	and Besseling, R. and Hesselberth, M. B. S. and Kes, P. H.},
  title = {Influence of edge barriers on vortex dynamics in thin weak-pinning
	superconducting strips},
  journal = {Phys. Rev. B},
  year = {2001},
  volume = {64},
  pages = {014503},
  month = {Jun},
  doi = {10.1103/PhysRevB.64.014503},
  issue = {1},
  numpages = {6},
  publisher = {American Physical Society},
  url = {http://link.aps.org/doi/10.1103/PhysRevB.64.014503}
}

@ARTICLE{Bez22prb,
  author = {Bezuglyj, A. I. and Shklovskij, V. A. and Budinsk\'a, B. and Aichner,
	B. and Bevz, V. M. and Mikhailov, M. Yu. and Vodolazov, D. Yu. and
	Lang, W. and Dobrovolskiy, O. V.},
  title = {Vortex jets generated by edge defects in current-carrying superconductor
	thin strips},
  journal = {Phys. Rev. B},
  year = {2022},
  volume = {105},
  pages = {214507},
  doi = {10.1103/PhysRevB.105.214507},
  issue = {21},
  numpages = {12},
  publisher = {American Physical Society},
  url = {https://link.aps.org/doi/10.1103/PhysRevB.105.214507}
}

@article{Gut09prb,
  title = {Transition from turbulent to nearly laminar vortex flow in superconductors with periodic pinning},
  author = {Gutierrez, J. and Silhanek, A. V. and Van de Vondel, J. and Gillijns, W. and Moshchalkov, V. V.},
  journal = {Phys. Rev. B},
  volume = {80},
  issue = {14},
  pages = {140514},
  numpages = {4},
  year = {2009},
  month = {Oct},
  publisher = {American Physical Society},
  doi = {10.1103/PhysRevB.80.140514},
  url = {https://link.aps.org/doi/10.1103/PhysRevB.80.140514}
}

@ARTICLE{Bud22pra,
  author = {Budinsk\'a, B. and Aichner, B. and Vodolazov, D. Yu. and Mikhailov,
	M. Yu. and Porrati, F. and Huth, M. and Chumak, A.V. and Lang, W.
	and Dobrovolskiy, O.V.},
  title = {Rising Speed Limits for Fluxons via Edge-Quality Improvement in Wide
	{MoSi} Thin Films},
  journal = {Phys. Rev. Appl.},
  year = {2022},
  volume = {17},
  pages = {034072},
  doi = {10.1103/PhysRevApplied.17.034072},
  issue = {3},
  numpages = {12},
  publisher = {American Physical Society},
  url = {https://link.aps.org/doi/10.1103/PhysRevApplied.17.034072}
}

@ARTICLE{Bra95rpp,
  author = {Brandt, E. H.},
  title = {The flux-line lattice in superconductors},
  journal = {Rep. Progr. Phys.},
  year = {1995},
  volume = {58},
  pages = {1465},
  url = {http://stacks.iop.org/0034-4885/58/i=11/a=003}
}

@ARTICLE{Dob26sst,
author = {Dobrovolskiy, O. and Suderow, H. and Tafuri, F. and Black-Schaffer, A. M. and 
Lado, J. L. and Sudb\o{}, A. and Stornaioulo, D. and Li, C. and B\"ohmer, A. E. and 
Tran, L. M. and Zaleski, A. J. and Crisan, A. and Polichetti, M. and Galluzzi, A. and 
Gencer, A. and Aichner, B. and Bari\'si\'c, N. and Lang, W. and Samuely, T. and 
Gmitra, M. and Cren, T. and Calandra, M. and Samuely, P. and Custers, J. and 
C\'ordoba, R. and Fomin, V. M. and Poccia, N. and Szab\'o, P. and Porrati, F. and 
Kakazei, G. and Aarts, J. and Robinson, J. and Villegas, J. E. and 
Althammer, M. and Huebl, H. and Kamra, A. and Weiler, M. and Dil, J. H. and 
Evtushinsky, D. and Kalisky, B. and Anahory, Y. and Bending, S. and 
Liljeroth, P. and Hassanien, A. and Guillam\'on, I. and Herrera, E. and 
Silhanek, A. V. and Van de Vondel, J. and Palau, A. and Charaev, I. and 
Sidorova, M. and Lombardi, F. and Bauch, T. and Feuillet-Palma, C. and 
Stolyarov, V. and Roditchev, D. and Krasnov, V. M. and Hampel, B. and 
Mart\'{\i}nez-P\'erez, M. J. and Ses\'e, J. and Koelle, D. and Poletto, S. and 
Bruno, A. and Massarotti, D.
},  
title = {Roadmap on nanoscale superconductivity for quantum technologies},
  journal = {Supercond. Sci. Technol.},
  year = {2026},
  volume = {39},
  pages = {023502},
  number = {2},
  doi = {10.1088/1361-6668/ae3030},
  publisher = {IOP Publishing},
  url = {https://doi.org/10.1088/1361-6668/ae3030}
}

@article{Bis23cpc,
title = {{pyTDGL}: {Time-dependent Ginzburg-Landau} in {Python}},
author = {Bishop-Van Horn, Logan},
journal2 = {Computer Physics Communications},
journal = {Comput. Phys. Commun.},
isbn = {0010-4655},
volume = {291},
year = {2023},
pages = {108799},
doi = {10.1016/j.cpc.2023.108799},

date = {2023/10/01/}
}

@ARTICLE{Rui26pms,
author = {
Ruiz, H. S. and Hänisch, J. and Polichetti, M. and Galluzzi, A. and 
Gozzelino, L. and Torsello, D. and Milo{\v s}evi\'c-Govedarovi\'c, S. and 
Grbovi\'c-Novakovi\'c, J. and Dobrovolskiy, O. V. and Lang, W. and 
Grimaldi, G. and Crisan, A. and Badica, P. and Ionescu, A. M. and 
Cayado, P. and Willa, R. and Barbiellini, B. and Eley, S. and 
Bad\'{\i}a-Maj\'{o}s, A.
},
  title = {Critical current density in advanced superconductors},
  journal = {Progr. Mater. Sci.},
  year = {2026},
  volume = {155},
  pages = {101492},
  doi = {https://doi.org/10.1016/j.pmatsci.2025.101492},
  issn = {0079-6425},
  url = {https://www.sciencedirect.com/science/article/pii/S0079642525000702}
}

@ARTICLE{Pat21prb,
  author = {Pathirana, W. P. M. R. and Gurevich, A.},
  title = {Effect of random pinning on nonlinear dynamics and dissipation of
	a vortex driven by a strong microwave current},
  journal = {Phys. Rev. B},
  year = {2021},
  volume = {103},
  pages = {184518},
  month = {May},
  doi = {10.1103/PhysRevB.103.184518},
  issue = {18},
  numpages = {19},
  publisher = {American Physical Society},
  url = {https://link.aps.org/doi/10.1103/PhysRevB.103.184518}
}

@ARTICLE{Vod19sst,
  author = {Vodolazov, D. Yu.},
  title = {Flux-flow instability in a strongly disordered superconducting strip
	with an edge barrier for vortex entry},
  journal = {Supercond. Sci. Technol.},
  year = {2019},
  volume = {32},
  pages = {115013},
  number = {11},
  month = {oct},
  doi = {10.1088/1361-6668/ab4168},
  publisher = {{IOP} Publishing},
  url = {https://doi.org/10.1088/1361-6668/ab4168}
}

@ARTICLE{Kar24pra,
author = {
Karrer, M. and Aichner, B. and Wurster, K. and Mag\'en, C. and 
Schmid, C. and Hutt, R. and Budinsk\'a, B. and 
Dobrovolskiy, O. V. and Kleiner, R. and Lang, W. and 
Goldobin, E. and Koelle, D.
},
  title = {Vortex matching at 6 {T} in {YBa}$_{2}${Cu}$_{3}${O}$_{7\ensuremath{-}\ensuremath{\delta}}$
	thin films by imprinting a 20-nm periodic pinning array with a focused
	helium-ion beam},
  journal = {Phys. Rev. Appl.},
  year = {2024},
  volume = {22},
  pages = {014043},
  month = {Jul},
  doi = {10.1103/PhysRevApplied.22.014043},
  issue = {1},
  numpages = {11},
  publisher = {American Physical Society},
  url = {https://link.aps.org/doi/10.1103/PhysRevApplied.22.014043}
}

@ARTICLE{Ada15prb,
  author = {Adami, O.-A. and Jelic, Z. L. and Xue, C. and Abdel-Hafiez, M. and
	Hackens, B. and Moshchalkov, V. V. and Milosevic, M. V. and Van de
	Vondel, J. and Silhanek, A. V.},
  title = {Onset, evolution, and magnetic braking of vortex lattice instabilities
	in nanostructured superconducting films},
  journal = {Phys. Rev. B},
  year = {2015},
  volume = {92},
  pages = {134506},
  month = {Oct},
  doi = {10.1103/PhysRevB.92.134506},
  issue = {13},
  numpages = {9},
  publisher = {American Physical Society},
  url = {https://link.aps.org/doi/10.1103/PhysRevB.92.134506}
}

@ARTICLE{Sil12njp,
  author = {Silhanek, A. V. and Leo, A. and Grimaldi, G. and Berdiyorov, G. R.
	and Milosevic, M. V and Nigro, A. and Pace, S. and Verellen, N. and
	Gillijns, W. and Metlushko, V. and Ili\'c, B. and Zhu, X. and Moshchalkov,
	V. V.},
  title = {Influence of artificial pinning on vortex lattice instability in
	superconducting films},
  journal = {New J. Phys.},
  year = {2012},
  volume = {14},
  pages = {053006},
  number = {5},
  url = {http://stacks.iop.org/1367-2630/14/i=5/a=053006}
}

@ARTICLE{Nag01nat,
  author = {Nagamatsu, J. and Nakagawa, N. and Muranaka, T. and
	Zenitani, Y. and Akimitsu, J.},
  title = {Superconductivity at {39 K} in magnesium diboride},
  journal = {Nature},
  year = {2001},
  volume = {410},
  pages = {63--64},
  number = {6824},
  doi = {10.1038/35065039},
  url = {https://doi.org/10.1038/35065039}
}

@ARTICLE{Mus80etp,
  author = {Musienko, L. E. and Dmitrenko, I. M. and Volotskaya, V. G.},
  title = {Nonlinear conductivity of thin films in a mixed state},
  journal = {JETP Lett.},
  year = {1980},
  volume = {31},
  pages = {567}
}

@Article{Dmi81phb,
author={Dmitrenko, I. M.
and Volotskaya, V. G.
and Musienko, L. E.
and Sivakov, A. G.},
title={Non-equilibrium effects in dynamic mixed state of thin films},
journal={Physica B+C},
year={1981},
month={Aug},
day={01},
volume={108},
number={1},
pages={783-784},
url={https://www.sciencedirect.com/science/article/pii/0378436381906963}
}

@article{Dmi06sst,
  author    = {Dmitriev, V. M.  and Zolochevskii, I. V.},
  title     = {Resistive current states in wide superconducting films in zero magnetic field},
  journal   = {Supercond. Sci. Technol.},
  year      = {2006},
  volume    = {19},
  number    = {4},
  pages     = {342--349},
  doi       = {10.1088/0953-2048/19/4/017},
  publisher = {IOP Publishing}
}

@PHDTHESIS{Gru24msc,
  author = {Markus Gruber},
  title = {Nonlinear conductivity and flux-flow instabilities in superconducting
	{MgB$_2$} thin films},
  school = {University of Vienns},
  year = {2024},
  type = {{MSc} Thesis}
}

@article{Sch26prb,
  author = {
    Schmid, Clemens and 
    Pokusinskyi, Anton and 
    Gruber, Markus and 
    Pfaff, Corentin and 
    Courtois, Theo and 
    Badie, Laurent and 
    Kasatkin, Alexander and 
    Dumesnil, Karine and 
    Mangin, Stephane and 
    Hauet, Thomas and 
    Dobrovolskiy, Oleksandr
    },
  title = {Crystal structure effects on vortex dynamics in superconducting {MgB}$_{2}$ thin films},
  journal = {Phys. Rev. B},
  pages = {},
  year = {2026},
  month = {Sep},
  publisher = {American Physical Society},
  doi = {10.1103/ghfm-2tzs},
  url = {https://link.aps.org/doi/10.1103/ghfm-2tzs}
}

@ARTICLE{Bev23pra,
  author = {Bevz, V.M. and Mikhailov, M.Yu. and Budinsk\'a, B. and Lamb-Camarena,
	S. and Shpilinska, S.O. and Chumak, A.V. and Urb\'anek, M. and Arndt,
	M. and Lang, W. and Dobrovolskiy, O.V.},
  title = {Vortex Counting and Velocimetry for Slitted Superconducting Thin
	Strips},
  journal = {Phys. Rev. Appl.},
  year = {2023},
  volume = {19},
  pages = {034098},
  doi = {10.1103/PhysRevApplied.19.034098},
  issue = {3},
  numpages = {14},
  url = {https://link.aps.org/doi/10.1103/PhysRevApplied.19.034098}
}

@ARTICLE{Kle85ltp,
  author = {Klein, W. and Huebener, R. P. and Gauss, S. and Parisi, J.},
  title = {Nonlinearity in the flux-flow behavior of thin-film superconductors},
  journal = {J. Low Temp. Phys.},
  year = {1985},
  volume = {61},
  pages = {413--432},
  number = {5},
  doi = {10.1007/BF00683694},
  issn = {1573-7357},
  url = {http://dx.doi.org/10.1007/BF00683694}
}

@ARTICLE{Tsu03prl,
  author = {Tsuda, S. and Yokoya, T. and Takano, Y. and Kito, H. and Matsushita,
	A. and Yin, F. and Harima, H. and Shin, S.},
  title = {Definitive experimental evidence for two-band superconductivity in
	{MgB}$_2$},
  journal = {Phys. Rev. Lett.},
  year = {2003},
  volume = {91},
  pages = {127001},
  number = {12},
  doi = {10.1103/PhysRevLett.91.127001},
  url = {https://doi.org/10.1103/PhysRevLett.91.127001}
}

@Article{Pfa24apl,
author={Pfaff, C.
and Petit-Watelot, S.
and Andrieu, S.
and Pasquier, L.
and Ghanbaja, J.
and Mangin, S.
and Dumesnil, K.
and Hauet, T.},
title={Spin injection at {MgB2}-superconductor/ferromagnet interface},
journal={Appl. Phys. Lett.},
year={2024},
month={Sep},
day={03},
volume={125},
number={10},
pages={102601},
doi={10.1063/5.0220815},
url={https://doi.org/10.1063/5.0220815}
}

@ARTICLE{Pok24ltp,
  author = {Pokusinskyi, A. O. and Kasatkin, A. L.},
  title = {Dissociation of composite Abrikosov vortices in two-band superconductors
	in a strong rf field},
  journal = {Low Temp. Phys.},
  year = {2024},
  volume = {50},
  pages = {111--116},
  number = {2},
  doi = {10.1063/10.0024321}
}

@ARTICLE{Pok25rrl,
  author = {Pokusinskyi, Anton O. and Dobrovolskiy, Oleksandr V.},
  title = {{DC}-Driven Fractional Flux Quanta in Two-Band Superconductors},
  journal = {Phys. Stat. Sol. -- Rap. Res. Lett.},
  year = {2025},
  volume = {19},
  pages = {2500128},
  number = {12},
  doi = {10.1002/pssr.202500128},
  url = {https://doi.org/10.1002/pssr.202500128}
}

@ARTICLE{Kom14apl,
  author = {Kompaniiets, M. and Dobrovolskiy, O. V. and Neetzel, C. and Porrati,
	F. and Br\"otz, J. and Ensinger, W. and Huth, M.},
  title = {Long-range superconducting proximity effect in polycrystalline {Co}
	nanowires},
  journal = {Appl. Phys. Lett.},
  year = {2014},
  volume = {104},
  pages = {052603},
  doi = {10.1063/1.4863980},
  url = {http://dx.doi.org/10.1063/1.4863980}
}

@ARTICLE{Pam94prb,
  author = {Pambianchi, M. S. and Mao, J. and Anlage, S. M.},
  title = {Magnetic screening in proximity-coupled superconductor/normal-metal
	bilayers},
  journal = {Phys. Rev. B},
  year = {1994},
  volume = {50},
  pages = {13659--13663},
  month = {Nov},
  doi = {10.1103/PhysRevB.50.13659},
  issue = {18},
  numpages = {0},
  publisher = {American Physical Society},
  url = {https://link.aps.org/doi/10.1103/PhysRevB.50.13659}
}

@ARTICLE{Per02pcs,
  author = {Peroz, C. and Villard, C. and Sulpice, A. and Butaud, P.},
  title = {Vortex dynamics at high velocities and proximity effect in superconducting
	thin films},
  journal = {Physica C},
  year = {2002},
  volume = {369},
  pages = {222-226},
  number = {1},
  issn = {0921-4534},
  url = {https://www.sciencedirect.com/science/article/pii/S0921453401012461}
}

@ARTICLE{Bez84ltp,
  author = {Bezuglyj, A. I. and Shklovskij, V. A.},
  title = {Thermal domains in inhomogeneous current-carrying superconductors.
	{C}urrent-voltage characteristics and dynamics of domain formation
	after current jumps},
  journal = {J. Low Temp. Phys.},
  year = {1984},
  volume = {57},
  pages = {227--247},
  number = {3},
  month = {Nov},
  day = {01},
  doi = {10.1007/BF00681190},
  issn = {1573-7357},
  url = {https://doi.org/10.1007/BF00681190}
}

@ARTICLE{Sil10pcs,
  author = {Silhanek, A. V. and Kramer, R.G.B. and Van de Vondel, J. and Moshchalkov,
	V.V. and Milosevic, M.V. and Berdiyorov, G.R. and Peeters, F.M. and
	Luccas, R.F. and Puig, T.},
  title = {Freezing vortex rivers},
  journal = {Physica C},
  year = {2010},
  volume = {470},
  pages = {726--729},
  number = {19},
  doi = {https://doi.org/10.1016/j.physc.2010.02.072},
  issn = {0921-4534},
  url = {http://www.sciencedirect.com/science/article/pii/S0921453410001796}
}

@ARTICLE{Siv03prl,
  author = {Sivakov, A. G. and Glukhov, A. M. and Omelyanchouk, A. N. and Koval,
	Y. and M\"uller, P. and Ustinov, A. V.},
  title = {Josephson Behavior of Phase-Slip Lines in Wide Superconducting Strips},
  journal = {Phys. Rev. Lett.},
  year = {2003},
  volume = {91},
  pages = {267001},
  doi = {10.1103/PhysRevLett.91.267001},
  issue = {26},
  numpages = {4},
  publisher = {American Physical Society},
  url = {http://link.aps.org/doi/10.1103/PhysRevLett.91.267001}
}

@ARTICLE{Att12pcm,
  author = {Attanasio, C. and Cirillo, C.},
  title = {Quasiparticle relaxation mechanisms in superconductor/ferromagnet
	bilayers},
  journal = {J. Phys.: Cond. Matt.},
  year = {2012},
  volume = {24},
  pages = {083201},
  number = {8},
  url = {http://stacks.iop.org/0953-8984/24/i=8/a=083201}
}

@ARTICLE{Ust24prb,
  author = {Ustavcshikov, S. S. and Levichev, M. Yu. and Pashenkin, I. Yu. and
	Gusev, N. S. and Mazilkin, A. A. and Vodolazov, D. Yu.},
  title = {Negative differential resistance and quasiperiodic vortex-antivortex
	motion in a superconducting constriction},
  journal = {Phys. Rev. B},
  year = {2024},
  volume = {109},
  pages = {174521},
  month = {May},
  doi = {10.1103/PhysRevB.109.174521},
  issue = {17},
  numpages = {8},
  publisher = {American Physical Society},
  url = {https://link.aps.org/doi/10.1103/PhysRevB.109.174521}
}

@ARTICLE{Bez19prb,
  author = {Bezuglyj, A. I. and Shklovskij, V. A. and Vovk, R. V. and Bevz, V.
	M. and Huth, M. and Dobrovolskiy, O. V.},
  title = {Local flux-flow instability in superconducting films near {${T}_{c}$}},
  journal = {Phys. Rev. B},
  year = {2019},
  volume = {99},
  pages = {174518},
  doi = {10.1103/PhysRevB.99.174518},
  numpages = {9},
  publisher = {American Physical Society},
  url = {https://link.aps.org/doi/10.1103/PhysRevB.99.174518}
}

@ARTICLE{Emb17nac,
  author = {Embon, L. and Anahory, Y. and Jelic, Z. L. and Lachman, E. O. and
	Myasoedov, Y. and Huber, M. E. and Mikitik, G. P. and Silhanek, A.
	V. and Milosevic, M. V. and Gurevich, A. and Zeldov, E.},
  title = {Imaging of super-fast dynamics and flow instabilities of superconducting
	vortices},
  journal = {Nat. Commun.},
  year = {2017},
  volume = {8},
  pages = {85},
  doi = {10.1038/s41467-017-00089-3},
  issn = {2041-1723},
  url = {https://doi.org/10.1038/s41467-017-00089-3}
}

@ARTICLE{Mis07prb,
  author = {Misko, V. R. and Savel'ev, S. and Rakhmanov, A. L. and Nori, F.},
  title = {Negative differential resistivity in superconductors with periodic
	arrays of pinning sites},
  journal = {Phys. Rev. B},
  year = {2007},
  volume = {75},
  pages = {024509},
  month = {Jan},
  doi = {10.1103/PhysRevB.75.024509},
  issue = {2},
  numpages = {10},
  publisher = {American Physical Society},
  url = {https://link.aps.org/doi/10.1103/PhysRevB.75.024509}
}

@ARTICLE{Kun02prl,
  author = {Kunchur, M. N.},
  title = {Unstable Flux Flow due to Heated Electrons in Superconducting Films},
  journal = {Phys. Rev. Lett.},
  year = {2002},
  volume = {89},
  pages = {137005},
  month = {Sep},
  doi = {10.1103/PhysRevLett.89.137005},
  issue = {13},
  numpages = {4},
  publisher = {American Physical Society},
  url = {http://link.aps.org/doi/10.1103/PhysRevLett.89.137005}
}

@ARTICLE{Kun01prl,
  author = {Kunchur, M. N. and Ivlev, B. I. and Knight, J. M.},
  title = {Steps in the Negative-Differential-Conductivity Regime of a Superconductor},
  journal = {Phys. Rev. Lett.},
  year = {2001},
  volume = {87},
  pages = {177001},
  month = {Oct},
  doi = {10.1103/PhysRevLett.87.177001},
  issue = {17},
  numpages = {4},
  publisher = {American Physical Society},
  url = {https://link.aps.org/doi/10.1103/PhysRevLett.87.177001}
}

@ARTICLE{Lar76etp,
  author = {Larkin, A. I. and Ovchinnikov, Yu. N.},
  title = {Nonlinear Conductivity of Superconductors in the Mixed State},
  journal = {Sov. Phys. JETP},
  year = {1976},
  volume = {41},
  pages = {960}
}

@ARTICLE{Kog22prb,
  author = {Kogan, V. G. and Nakagawa, N.},
  title = {Dissipation of moving vortices in thin films},
  journal = {Phys. Rev. B},
  year = {2022},
  volume = {105},
  pages = {L020507},
  month = {Jan},
  doi = {10.1103/PhysRevB.105.L020507},
  issue = {2},
  numpages = {4},
  publisher = {American Physical Society},
  url = {https://link.aps.org/doi/10.1103/PhysRevB.105.L020507}
}

@ARTICLE{Kog20prb,
  author = {Kogan, V. G. and Prozorov, R.},
  title = {Interaction between moving {Abrikosov} vortices in {type-II} superconductors},
  journal = {Phys. Rev. B},
  year = {2020},
  volume = {102},
  pages = {024506},
  month = {Jul},
  doi = {10.1103/PhysRevB.102.024506},
  issue = {2},
  numpages = {6},
  publisher = {American Physical Society},
  url = {https://link.aps.org/doi/10.1103/PhysRevB.102.024506}
}

@article{Kos16prb,
title = {Optimization of vortex pinning by nanoparticles using simulations of the time-dependent {G}inzburg-{L}andau model},
author = {Koshelev, A. E. and Sadovskyy, I. A. and Phillips, C. L. and Glatz, A.},
journal = {Phys. Rev. B},
volume = {93},
issue = {6},
pages = {060508},
numpages = {5},
year = {2016},
month = {Feb},
publisher = {American Physical Society},
doi = {10.1103/PhysRevB.93.060508},
url = {https://link.aps.org/doi/10.1103/PhysRevB.93.060508}
}

@dataset{Pok26mendeley,
  author    = {Pokusinskyi, Anton},
  title     = {Vortex rivers and multiple voltage transitions in superconducting {MgB2} thin films},
  year      = {2026},
  publisher = {Mendeley Data},
  version   = {1},
  doi       = {10.17632/8bghjwyrts.1}
}
\end{document}